\documentclass[twocolumn,showpacs,aps,prd,floatfix]{revtex4}
\usepackage{graphicx}
\usepackage{dcolumn}
\usepackage{amsmath}
\usepackage{epsfig}

\newcommand{\BaBarType}      {PUB}  
\newcommand{\BaBarYear}       {05}
\newcommand{\BaBarNumber}     {12}
\newcommand{\SLACPubNumber} {11137}

\def\sss{\scriptscriptstyle}
\def\barpd{{\raise.35ex\hbox
{${\sss (}$}}--{\raise.35ex\hbox{${\sss )}$}}}
\def\dbarp{\tilde{D}^0}

\def\dbarpstar{\tilde{D}^{*0}}
\def\Dstarz  {\ensuremath{D^{*0}}\xspace}
\def\Dstarm  {\ensuremath{D^{*-}}\xspace}

\def\Kstarz  {\ensuremath{K^{*0}}\xspace}

\def\Kstarp  {\ensuremath{K^{*+}}\xspace}

\def\Kpm      {\ensuremath{K^{\pm}}\xspace}
\def\Dz      {\ensuremath{D^0}\xspace}
\def\piz   {\ensuremath{\pi^{0}}\xspace}
\def\pim   {\ensuremath{\pi^{-}}\xspace}
\def\pip   {\ensuremath{\pi^{+}}\xspace}
\def\KS    {\ensuremath{K_{\scriptscriptstyle S}}\xspace}

\def\mes        {\mbox{$m_{\rm ES}$}\xspace}

\def\to         {\ensuremath{\rightarrow}\xspace}

\def\Dzbpar  {\ensuremath{\Dbar^{(*)0}}\xspace}

\input pubboard/babarsym

\long\def\inst#1{\par\nobreak\kern 4pt\nobreak
    {\it #1}\par\vskip 10pt plus 3pt minus 3pt}

\begin{document}

\begin{flushleft}
\babar-\BaBarType-\BaBarYear/\BaBarNumber \\
SLAC-PUB-\SLACPubNumber \\
\end{flushleft}

\title{Search for {\boldmath $b \rightarrow u$} transitions in 
{\boldmath $B^- \to D^0 K^-$} and 
{\boldmath $B^- \to D^{*0} K^-$}}

%
\author{B.~Aubert}
\author{R.~Barate}
\author{D.~Boutigny}
\author{F.~Couderc}
\author{Y.~Karyotakis}
\author{J.~P.~Lees}
\author{V.~Poireau}
\author{V.~Tisserand}
\author{A.~Zghiche}
\affiliation{Laboratoire de Physique des Particules, F-74941 Annecy-le-Vieux, France }
\author{E.~Grauges}
\affiliation{IFAE, Universitat Autonoma de Barcelona, E-08193 Bellaterra, Barcelona, Spain }
% ...
\collaboration{The \babar\ Collaboration}
\noaffiliation

\begin{abstract}
\noindent
We search for $B^- \to \tilde{D}^0 K^-$ and $B^- \to
\tilde{D}^{*0} K^-$ and charge conjugates.
Here the symbol $\tilde{D}^0$ indicates 
decay of a $D^0$ or $\bar{D}^0$ into $K^+\pi^-$,
while the symbol $\tilde{D}^{*0}$ indicates
decay of a $D^{*0}$ or $\bar{D}^{*0}$ to
$\tilde{D}^0 \pi^0$ or 
$\tilde{D}^0 \gamma$.
These final states
can be reached through the $b \to c$
transition $B^- \to D^{(*)0}K^-$ followed by the doubly
Cabibbo-Kobayashi-Maskawa
(CKM)-suppressed $D^0 \to K^+ \pi^-$, or the $b \to u$
transition $B^- \to \Dzbpar K^-$ followed by the CKM-favored $\Dzb
\to K^+ \pi^-$.
The interference of these two amplitudes
is sensitive to the angle $\gamma$ of the unitarity triangle.
Our results are
based on 232 million $\FourS \to B\Bbar$ decays collected with the
\babar\ detector at SLAC.   We find no significant evidence
for these decays.
We set a limit 
$r_B \equiv |A(B^- \to \Dzb K^-)/A(B^- \to \Dz K^-)| < 0.23$ at  
90\% C.L. using the most conservative 
assumptions on the values of the CKM angle $\gamma$ and
the strong phases in the $B$ and $D$ decay amplitudes.  
In the case of the $D^*$ we set a 90\% C.L. limit
$r^{*2}_B \equiv |A(B^- \to \Dstarzb K^-)/A(B^- \to \Dstarz K^-)|^2 
< (0.16)^2$ which is independent of assumptions on $\gamma$ 
and strong phases.

\end{abstract}

\maketitle

\section{INTRODUCTION}
\label{sec:Introduction}
   Following the discovery of \CP violation in $B$-meson 
   decays and the measurement of the angle $\beta$
   of the unitarity triangle~\cite{cpv} associated with
   the Cabibbo-Kobayashi-Maskawa (CKM) quark mixing matrix, focus has turned
   towards the measurements of the other angles $\alpha$ and $\gamma$.
   The angle $\gamma$ is ${\rm arg}(-V_{ub}^*V^{}_{ud}/V_{cb}^*V^{}_{cd})$,
   where $V^{}_{ij}$ are CKM matrix elements.
   In the Wolfenstein convention~\cite{wolfenstein},
   $\gamma = {\rm arg}(V_{ub}^*)$.

\begin{figure}[hb]
   \begin{center}
      \epsfig{file=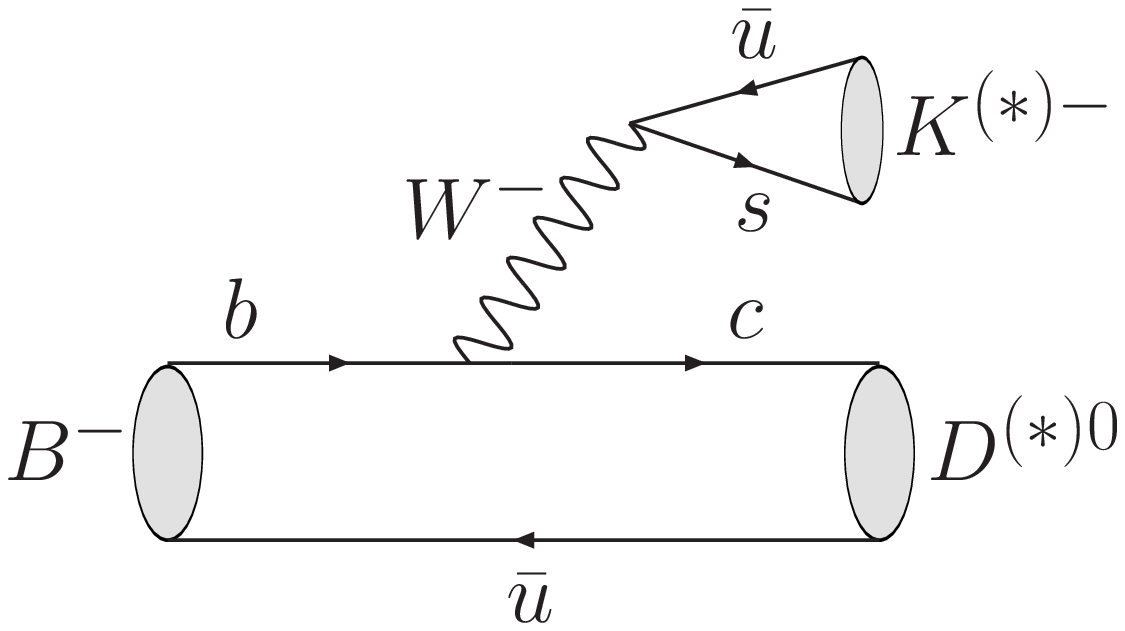,width=0.47\linewidth}
      \epsfig{file=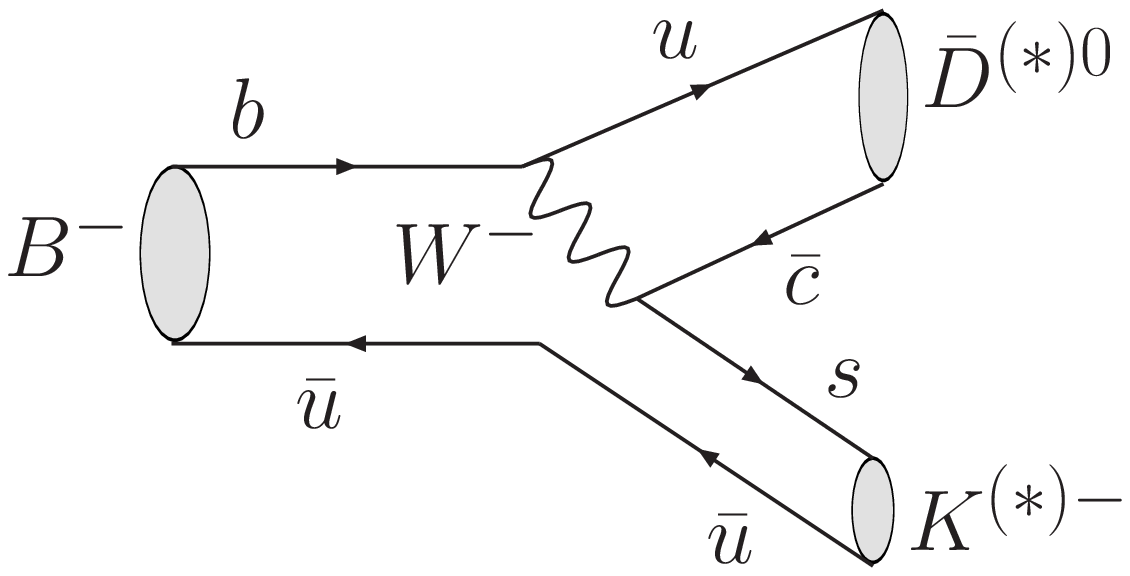,width=0.47\linewidth}
   \caption{Feynman diagrams for $B^- \to D^{(*)0} K^{(*)-}$ and $\Dzbpar K^{(*)-}$.
   The latter is CKM and color suppressed with respect to the former.
   The CKM-suppression factor is 
   $|V_{ub}V_{cs}^*/V_{cb}V_{us}^*| \approx 0.4$.  The naive
   color-suppression factor is $\frac{1}{3}$.}
   \label{fig:feynman}
   \end{center}
   \end{figure}

   Several proposed methods for measuring $\gamma$ exploit the
   interference between $B^- \to D^{(*)0}K^{(*)-}$ 
   and $B^- \to \Dzbpar K^{(*)-}$ 
   (Fig.~\ref{fig:feynman}) that occurs when the $D^{(*)0}$ and the
   \Dzbpar decay to common final states, as first suggested in Ref.~\cite{dk1}.

   As proposed in Ref.~\cite{dk2}, we search for $B^- \to \dbarp K^-$
   and $B^- \to \dbarpstar K^-,~ \dbarpstar \to \dbarp \pi^0$
   or $\dbarpstar \to \dbarp \gamma$,
   followed by $\dbarp \to K^+\pi^-$, as well as the charge conjugate (c.c.)
   sequences.  Here the symbol $\dbarp$ indicates the decay of a
   $\Dz$ or $\Dzb$ into $K^+ \pi^-$. 
   In these processes, the favored $B$ decay ($B^- \to D^{(*)0} K^-$)
   followed by
   the doubly CKM-suppressed $D$ decay ($\Dz \to K^+ \pi^-$)
   interferes with the suppressed
   $B$ decay ($B^- \to \Db^{(*)0} K^-$)
   followed by the CKM-favored $D$ decay ($\Dzb \to K^+ \pi^-$).
   The interference
   of the $b \to c$ ($B^- \to D^{(*)0} K^-$)
   and $b \to u$ ($B^- \to \Db^{(*)0} K^-$)
   amplitudes is sensitive to the
   relative weak phase $\gamma$.

   We use the notation $B^- \to [h_1^+h_2^-]_D h_3^-$
   (with each $h_i=\pi$ or $K$) for the decay chain $B^- \to \dbarp~h_3^-$,
   $\dbarp~\to h_1^+ h_2^-$.  For the closely related modes with a 
   $\dbarpstar$, we use the same notation with the subscript $D$ replaced
   by $D\piz$ or $D\gamma$, depending on whether the $\dbarpstar$
   decays to $\dbarp \piz$ or $\dbarp \gamma$.
   We also refer to $h_3$ as the bachelor
   $\pi$ or $K$.  

   In the decays of interest, the sign of the bachelor kaon is opposite to
   that of the kaon from $\dbarp$ decay.  It is convenient to
   define ratios of rates between these decays and the similar decays where
   the two kaons have the same sign.  The decays with same-sign kaons have much higher rate and
   proceed almost exclusively through the
   CKM-favored and color-favored $B$ transition, followed by the CKM-favored
   $D$-decay, {\it e.g.}, $B^- \to D^0 K^-$, $D^0 \to K^- \pi^+$.
   The advantage of taking ratios is that most
   theoretical and experimental uncertainties cancel. 
   Thus, ignoring possible small effects due to $D$ mixing
   and taking into account the effective phase difference of $\pi$ between
   the \Dstar decays in $D \gamma$ and $D \piz$~\cite{Bondar}, 
   we define the charge-specific ratios for $D$ and $D^*$ as:

   \begin{equation}
   {\cal R}^{\pm}_{K\pi} \equiv 
   \frac{\Gamma([K^\mp \pi^\pm]_D K^\pm)}{\Gamma([K^\pm \pi^\mp]_D K^\pm)}
   = r_B^2 + r_D^2 + 2 r_B r_D \cos(\pm \gamma + \delta),
   \end{equation}
   %
   %
   \begin{eqnarray}
   {\cal R}^{*\pm}_{K\pi,D\piz}  & \equiv & 
   \frac{\Gamma([K^\mp \pi^\pm]_{D\piz} K^\pm)}{\Gamma([K^\pm \pi^\mp]_{D\piz} K^\pm)} \nonumber 
\\
 &  =& r_B^{*2} + r_D^2 + 2 r_B^* r_D \cos(\pm \gamma + \delta^*), 
   \end{eqnarray}
\noindent and
   \begin{eqnarray}
   {\cal R}^{*\pm}_{K\pi,D\gamma}  & \equiv & 
   \frac{\Gamma([K^\mp \pi^\pm]_{D\gamma} K^\pm)}{\Gamma([K^\pm \pi^\mp]_{D\gamma} K^\pm)} \nonumber 
\\
&   =& r_B^{*2} + r_D^2 - 2 r_B^* r_D \cos(\pm \gamma + \delta^*),
   \end{eqnarray}
   \noindent where
   \begin{equation}
   r_B \equiv \left| \frac{A(B^- \to \Dzb K^-)}{A(B^- \to D^0 K^-)} \right|, 
   \end{equation}
   \begin{equation}
   r_B^* \equiv \left| \frac{A(B^- \to \Dstarzb K^-)}{A(B^- \to \Dstarz K^-)} \right|, 
   \end{equation}
   \begin{equation}
   \label{eqn:rd}
   r_D \equiv \left| \frac{A(D^0 \to K^+ \pi^-)}{A(D^0 \to K^- \pi^+)} \right|,
   \end{equation}
   \begin{equation}
   \delta^{(*)} \equiv \delta_B^{(*)} + \delta_D,
   \end{equation}
   and \noindent $\delta_B^{(*)}$ and $\delta_D$ are 
   strong phase differences between the two $B$ and $D$ decay 
   amplitudes, respectively.  The value of $r_D$ has been
   measured to be $r_D = 0.060 \pm 0.002$~\cite{PDG}.

   We also define the charge-independent ratio
   \begin{equation}
   {\cal R}_{K\pi} \equiv 
   \frac{\Gamma([K^+ \pi^-]_D K^-)+\Gamma([K^- \pi^+]_D K^+)}
        {\Gamma([K^- \pi^+]_D K^-)+\Gamma([K^+ \pi^-]_D K^+)}
   \end{equation}
   and the equivalent ratios for the $D^*$ modes,
   \begin{equation}
   {\cal R}_{K\pi,D\piz} \equiv 
   \frac{\Gamma([K^+ \pi^-]_{D\piz} K^-)+\Gamma([K^- \pi^+]_{D\piz} K^+)}
        {\Gamma([K^- \pi^+]_{D\piz} K^-)+\Gamma([K^+ \pi^-]_{D\piz} K^+)},
   \end{equation}
   and
   \begin{equation}
   {\cal R}_{K\pi,D\gamma} \equiv 
   \frac{\Gamma([K^+ \pi^-]_{D\gamma} K^-)+\Gamma([K^- \pi^+]_{D\gamma} K^+)}
        {\Gamma([K^- \pi^+]_{D\gamma} K^-)+\Gamma([K^+ \pi^-]_{D\gamma} K^+)}.
   \end{equation}
   Then,
   \begin{equation}
   {\cal R}_{K\pi} = \frac{{\cal R}^{+}_{K\pi}+{\cal R}^{-}_{K\pi}}{2} =
   r_B^{2} + r_D^2 + 2 r_B r_D \cos\gamma \cos \delta
   \label{eq:nostar}
   \end{equation}
   %
    and, similarly for the $D^*$ modes, 
   \begin{equation}
   {\cal R}^{*}_{K\pi,D\piz } = 
   r_B^{*2} + r_D^2 + 2 r_B^{*} r_D \cos\gamma \cos \delta^{*}, 
   \label{eq:star-pi}
   \end{equation}
   \begin{equation}
   {\cal R}^{*}_{K\pi,D\gamma} = 
   r_B^{*2} + r_D^2 - 2 r_B^{*} r_D \cos\gamma \cos \delta^{*}.
   \label{eq:star-gam}
   \end{equation}
   Equations~\ref{eq:nostar},~\ref{eq:star-pi}, and~\ref{eq:star-gam}
   assume no \CP violation in the normalization modes
   $[K^\mp \pi^\pm]_D K^\mp$, $[K^\mp \pi^\pm]_{D\piz} K^\mp$, and
   $[K^\mp \pi^\pm]_{D\gamma} K^\mp$.
   In the following we use the notation
   ${\cal R}^*_{K\pi}$ when there is no need to refer 
   specifically to ${\cal R}^*_{K\pi, D\piz}$
   or ${\cal R}^*_{K\pi, D\gamma}$.

   As discussed below, the parameter $r_B^{(*)}$ 
   is expected to be of the same order as $r_D$.  Thus, 
   \CP violation could manifest itself as a large difference between
   the charge-specific ratios
   ${\cal R}^{(*)+}_{K\pi}$ and ${\cal R}^{(*)-}_{K\pi}$.  Measurements of
   these six ratios can be used to constrain $\gamma$.

   The value of $r_B^{(*)}$ determines, in part, the level of interference
   between the diagrams
   of Fig.~\ref{fig:feynman}.  In most techniques for measuring $\gamma$,
   high values of $r_B^{(*)}$ lead to larger interference and better sensitivity to $\gamma$.
   Thus, $r_B$ and $r_B^{*}$ are key quantities whose values have a significant
   impact on the ability to measure the CKM angle $\gamma$ at the 
   $B$-factories and beyond.
%
%

   In the standard model,
   $r_B^{(*)} = |V^{}_{ub} V_{cs}^*/V^{}_{cb} V_{us}^*| \: F_{cs} \approx 0.4 \: F_{cs}$.
   The color-suppression factor 
   $F_{cs} < 1$ accounts for the additional suppression, beyond that
   due to CKM factors, of $B^- \to \Dzbpar K^-$ relative to $B^- \to D^{(*)0} K^-$.
   Naively, $F_{cs} = \frac{1}{3}$, which is the probability for the color
   of the quarks from the virtual $W$ in $B^- \to \Dzbpar K^-$ to match 
   that of the other two quarks; see Fig.~\ref{fig:feynman}.  
   Early estimates~\cite{neubert} 
   of $F_{cs}$ were based on factorization and the
   experimental information on a number of $b \to c$ transitions 
   available at the time.
   These estimates gave $F_{cs} \approx 0.22$, leading to
   $r_B^{(*)} \approx 0.09$.   However, the recent observations and
   measurements~\cite{colorsuppressed} 
   of color-suppressed $b \to c$ decays ($B \to D^{(*)} h^0$; 
   $h^0 = \pi^0, \rho^0, \omega, \eta, \eta'$) suggest 
  that color suppression is not as effective as anticipated,
  and therefore the value of $r_{B}^{(*)}$ could be of order 
  0.2~\cite{gronau}.

   As we will describe below, 
   the measured ${\cal R}^{(*)}_{K\pi}$ are consistent
   with zero in the current analysis.  
   Since ${\cal R}^{(*)}_{K\pi}$ depend quadratically on
   $r_B^{(*)}$, 
   we will use our measurements to set restrictive upper
   limits on $r_B^{(*)}$.

   It is important to note 
   the different signs of the third terms in the expressions
   for ${\cal R}^{*}_{K\pi,D\piz}$ and ${\cal R}^{*}_{K\pi,D\gamma}$
   in Equations~\ref{eq:star-pi} and~\ref{eq:star-gam}.
   This relative minus sign is due to the phase of $\pi$ between the 
   two $D^*$ decay modes~\cite{Bondar}.  It allows for a measurement of
   $r_B^*$ with no additional inputs since
   $r_B^{*2} = \frac{1}{2}( 
   {\cal R}^{*}_{D \piz}+{\cal R}^{*}_{D\gamma}) - r_D^2$,
   and $r_D$ is known quite precisely ($r_D = 0.060 \pm 0.002$).
   We will use this equation for $r_B^*$ and our results for 
   ${\cal R}^{*}_{D \piz}$ and ${\cal R}^{*}_{D\gamma}$
   to set an upper limit on $r_B^*$ with no assumptions.

   On the other hand, ${\cal R}_{K\pi}$ depends on the
   three unknowns $r_B$, $\gamma$, and $\delta$, see 
   Equation~\ref{eq:nostar}; thus, in order to extract a 
   limit on $r_B$ from the experimental limit 
   on ${\cal R}_{K\pi}$ we must make assumptions about
   $\gamma$ and $\delta$.  As we will discuss in 
   Section~\ref{sec:results}, we have chosen to quote
   an upper limit on $r_B$ based on the most conservative 
   assumptions on $\gamma$ and $\delta$.

   In this paper we report on an update of our previous search for  
   $B^- \to \dbarp K^-$~\cite{ADS-BABAR}, and the first attempt 
   to study $B^- \to \dbarpstar K^-$.  The previous analysis 
   was based on a sample of $B$-meson decays a factor of 1.9 smaller
   than used here,  and resulted in an upper limit ${\cal R}_{K\pi} < 0.026$ at the 90\% 
   C.L.  This in turn was translated into a limit $r_B < 0.22$,
   also at the 90\% confidence level. A similar analysis by the 
   Belle Collaboration~\cite{ADS-Belle} gives limits 
   ${\cal R}_{K\pi} < 0.044$ and $r_B < 0.27$ (90\% C.L.).

   Information on $r_B$, $r_B^*$, and $\gamma$ can also be obtained
   from studies 
   of $B^{\pm} \to \dbarp K^{\pm}$ and
   $B^{\pm} \to \dbarpstar K^{\pm}$, $\dbarp \to K_S \pi^+ \pi^-$.
   An analysis by the Belle collaboration~\cite{belle-Dalitz}
   finds quite large values 
   $r_B = 0.31 \pm 0.11$ and 
   $r_B^* = 0.34 \pm 0.14$, although the uncertainties are large enough
   that these results are not inconsistent with the limits listed above.
   On the other hand, a similar analysis by the \babar\ 
   Collaboration~\cite{babar-Dalitz}
   favors smaller values for these amplitude ratios:
   $r_B = 0.12 \pm 0.09$ at 90\% C.L. and $r^*_B = 0.17\pm 0.10$.

   \begin{table*}[tb]
   \caption{Notation used in the text for the decay modes that define the 
    data samples used in the analysis.}
   \begin{tabular*}{0.75\textwidth}{@{\extracolsep{\fill}}|l|l|l|} \hline
   Abbreviation & Mode & Comments\\ \hline
   $DK$      & $B^- \to D^0 K^-$, $D^0 \to K^- \pi^+$ and c.c. & normalization \\
   $D\pi$    & $B^- \to D^0 \pi^-$, $D^0 \to K^- \pi^+$ and c.c. & control  \\
   $\Dbar K$ & $B^- \to \dbarp K^-$, $\dbarp \to K^+\pi^-$ and c.c. & signal  \\
   $D^*K$    & $B^- \to D^{*0} K^-$, $D^{*0} \to D^0 \pi^0/\gamma$,  $D^0 \to K^- \pi^+$ and c.c. & normalization  \\
   $D^*\pi$    & $B^- \to D^{*0} \pi^-$, $D^{*0} \to D^0 \pi^0/\gamma$,  $D^0 \to K^- \pi^+$ and c.c. & control \\
   $\Dstarb K$& $B^- \to \dbarpstar~~K^-$, $\dbarpstar~~\to \dbarp \pi^0/\gamma$,
$\dbarp \to K^+ \pi^-$ and c.c. & signal  \\
   \hline
   \end{tabular*}
   \label{tab:shorthand}
   \end{table*}

\section{THE \babar\ DATASET}
\label{sec:babar}

   The results presented in this paper are based on 
    $232 \times 10^6$ $\FourS\to B\Bbar$ decays,
   corresponding to an integrated luminosity of 211 fb$^{-1}$.
   The data were collected
   between 1999 and 2004 with the \babar\ detector~\cite{babar} at the \pep2\
   \BF\ at SLAC~\cite{pep2}.  
   In addition, a 16~fb$^{-1}$ off-resonance data sample,
   with center-of-mass (CM) 
   energy 40~\mev below the \FourS resonance,
   is used to study backgrounds from
   continuum events, $e^+ e^- \to q \bar{q}$
   ($q=u,d,s,$ or $c$).

\section{ANALYSIS METHOD}
\label{sec:Analysis}
   This work is an extension of our 
   analysis from Ref.~\cite{ADS-BABAR}, which resulted in 
   90\% C.L. limits on
   ${\cal R}_{K\pi} < 0.026$ and $r_B < 0.22$, as mentioned 
   in Section~\ref{sec:Introduction}.
   The main changes in the analysis are
   the following:
   \begin{itemize}
   \item The size of the dataset is increased from 
   $120$ to $232 \times 10^6$ $\FourS\to B\Bbar$ decays.
   \item This analysis also includes the 
   $B^{\pm} \to \dbarpstar K^{\pm}$ mode.
   \item The event selection criteria have been 
   made more stringent 
   in order to 
   reduce backgrounds further.
   \item A few of the selection criteria
   in the previous analysis resulted
   in small differences in the efficiencies of the signal mode 
   $B^{\pm} \to [K^{\mp}\pi^{\pm}]_DK^{\pm}$ and the normalization mode
   $B^{\pm} \to [K^{\pm}\pi^{\mp}]_DK^{\pm}$.  These 
   selection criteria have now been removed.
   \end{itemize}

   \noindent The analysis makes use of several samples from different decay modes.  
   Throughout the following discussion we will refer to these modes 
   using abbreviations that
   are summarized in Table~\ref{tab:shorthand}.

   The event selection is developed from studies of
   simulated $B\Bbar$ and continuum events, and off-resonance
   data. A large on-resonance control sample of $D\pi$ and 
   $D^*\pi$ events 
   is used to validate several aspects of the simulation and 
   analysis procedure.

   The analysis strategy is the following:
   \begin{enumerate}
   \item The goal is to measure or set limits on the
   charge-independent ratios ${\cal R}_{K\pi}$ and ${\cal R}_{K\pi}^*$. 
   \item The first step consists of the application of a 
   set of basic selection criteria to select possible candidate events,
   see Section~\ref{sec:basic}.
   \item After the basic selection criteria, 
   the backgrounds are dominantly from
   continuum.  These are significantly reduced with a neural network designed
   to distinguish between $B\Bbar$ and continuum events,
   see Section~\ref{sec:nn}.
   \item After the neural network requirement, events are characterized by 
   two kinematical variables that are customarily used when reconstructing
   $B$-meson decays at the $\FourS$.  These variables are 
   the energy-substituted mass,
   $\mes \equiv 
   \sqrt{(\frac{s}{2}  + \vec{p}_0\cdot \vec{p}_B)^2/E_0^2 - p_B^2}$
   and energy difference $\Delta E \equiv E_B^*-\frac{1}{2}\sqrt{s}$, 
   where $E$ and $p$ are energy and momentum, the asterisk
   denotes the CM frame, the subscripts $0$ and $B$ refer to the
   \FourS and $B$ candidate, respectively, and $s$ is the square
   of the CM energy.  
   For signal events $\mes = m_B$ and
   $\Delta E = 0$ within the 
   resolution of about 2.5 and 20 MeV, respectively
   (here $m_B$ is the known $B$ mass~\cite{PDG}).
   \item We then perform simultaneous 
   unbinned maximum likelihood 
   fits to the final signal samples ($\Dbar K$ and $\Dstarb K$),
   the normalization samples ($DK$ and $\Dstar K$),
   and the control samples ($D\pi$ and $\Dstar \pi$)
   to extract ${\cal R}_{K\pi}$ and ${\cal R}_{K\pi}^*$,
   see Section~\ref{sec:fit}. 
   The fits are based on the reconstructed values of $\mes$ and 
   $\Delta E$ in the various event samples.
   \item Throughout the whole analysis chain, care is taken to 
   treat the signal, normalization, and control samples in a consistent manner.
   \end{enumerate}

    \begin{figure*} 
    \begin{center}
    \includegraphics[width=2.0in]{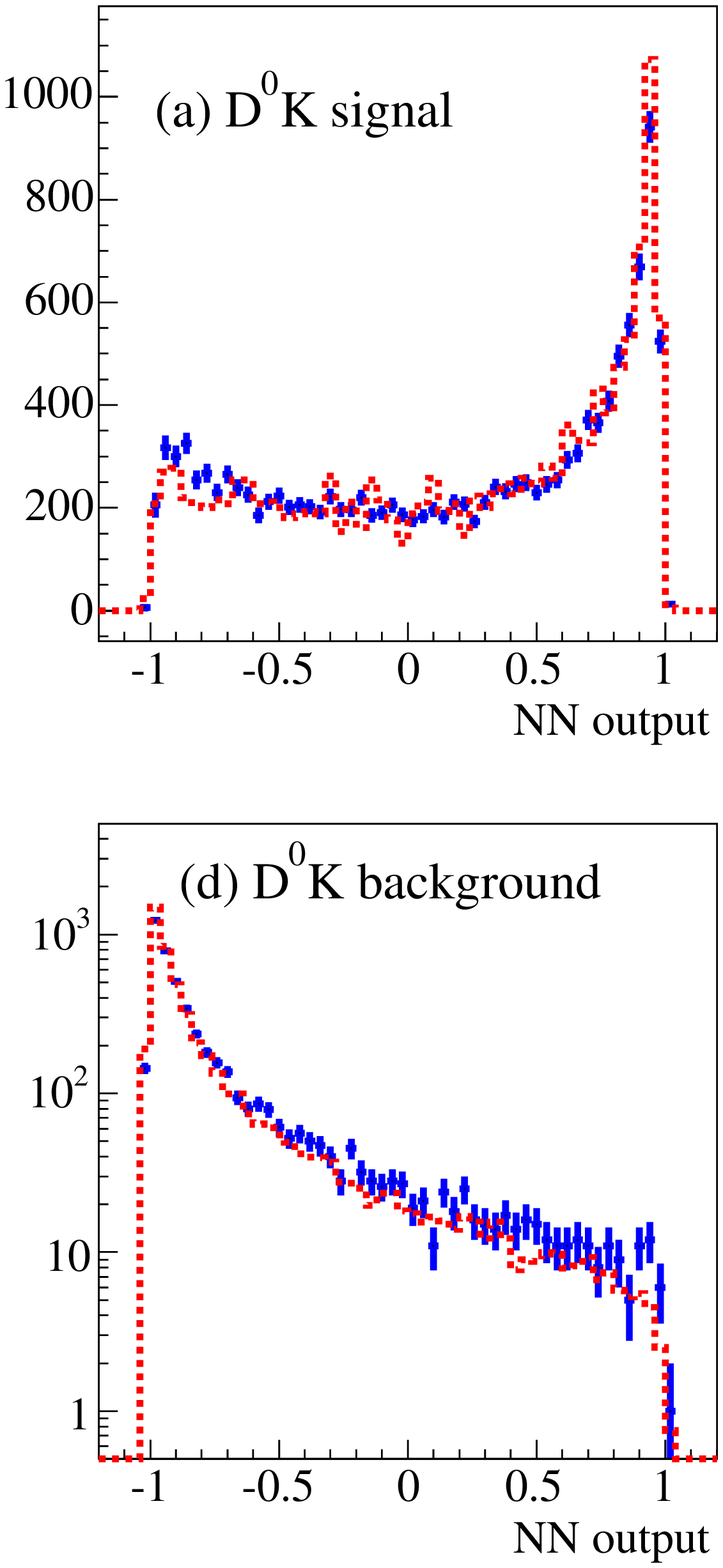}
    \includegraphics[width=2.0in]{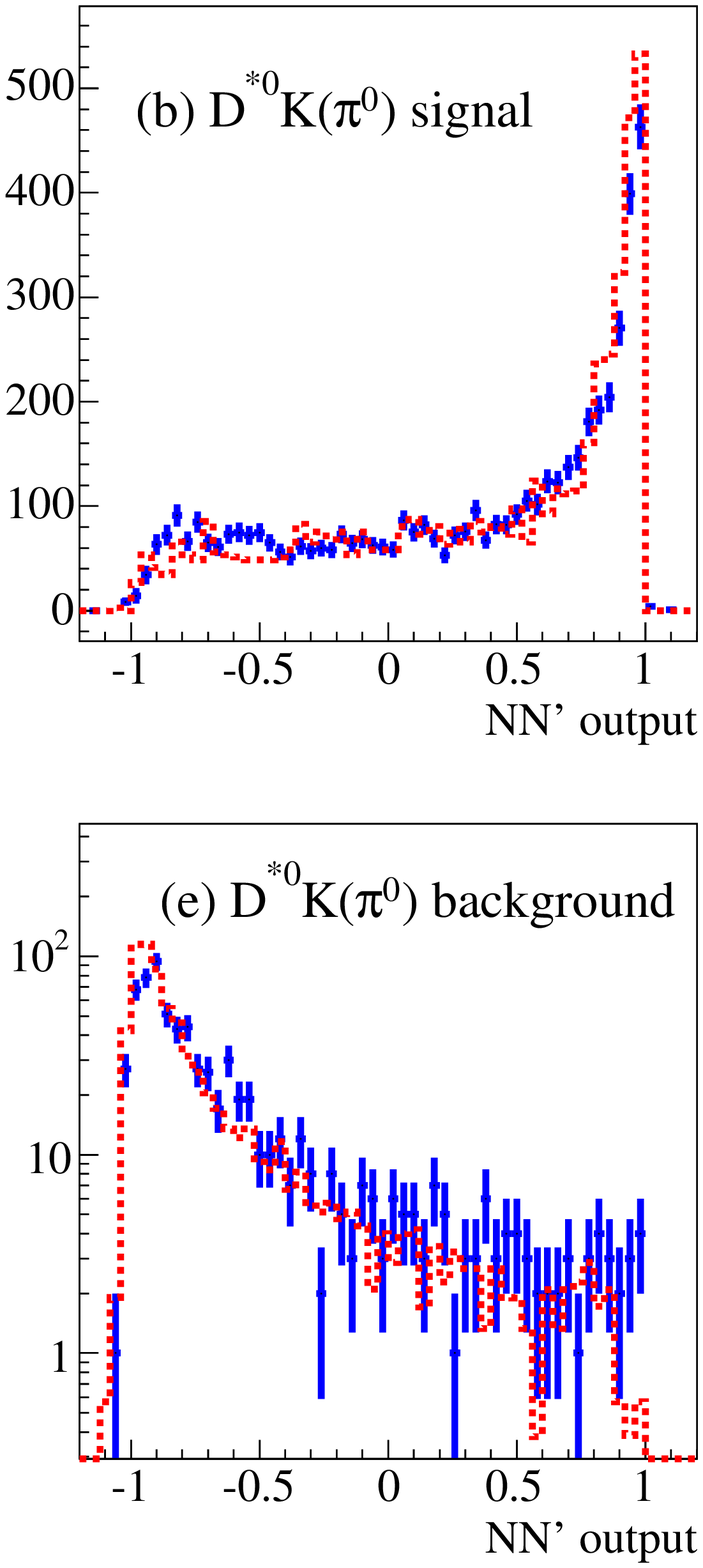}
    \includegraphics[width=2.0in]{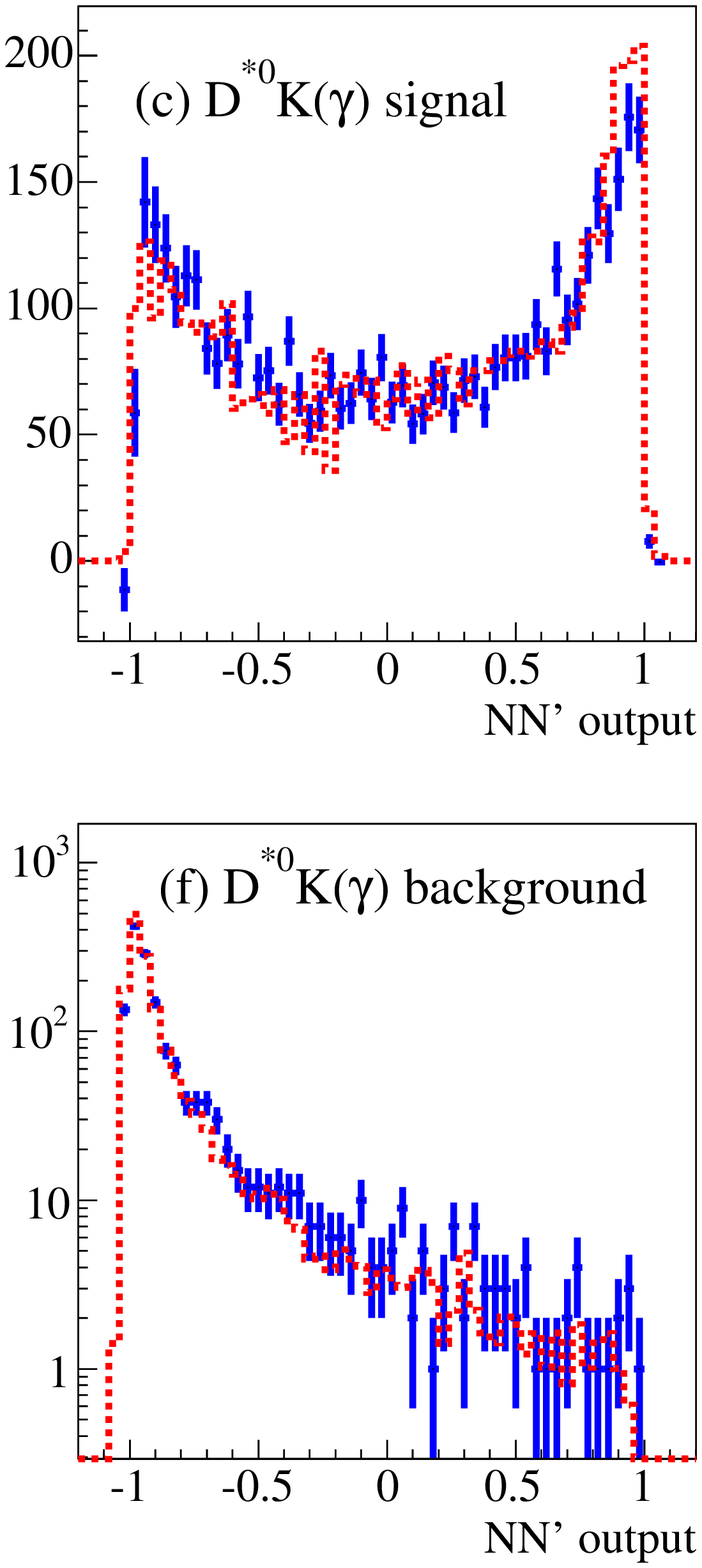}
    \caption{ Distributions of the continuum suppression neural
    network ($NN$ and $NN'$) outputs for the three modes. Figs.
    (a-c) show the expected distribution from signal events. The dashed
    line histogram shows the distribution of simulated signal events and
    the histogram with error bars shows the distribution of
    $D^{(*)0}\pi$ control sample events with background subtracted
    using the $\mes$ sideband (5.20 GeV $< \mes <$ 5.27 GeV). 
    Figs. (d-f) show the expected
    distribution for continuum background events. The dashed line
    histogram shows the distribution of simulated continuum events and
    the histogram with errors shows the distribution of off-resonance
    events. The $\mes$ and $\Delta E$ requirements on the
    off-resonance and continuum Monte Carlo events have been kept
    loose to increase the statistics. 
    Each Monte Carlo
    histogram is normalized to the area of the corresponding data
    histogram.}
    \label{fig:nnplots}
    \end{center}
    \end{figure*}

   \subsection{Basic Selection Criteria}
   \label{sec:basic}
   Charged kaon and pion candidates in the decay modes of interest
   must satisfy $K$ or $\pi$ identification criteria~\cite{PIDref}
   that are typically 85\% efficient, depending on momentum and polar angle.
   The misidentification rates are at the few percent level.

   The invariant mass of the $K\pi$ pair must be within 18.8 MeV
   (2.5$\sigma$) of the mean reconstructed $D^0$ mass (1863.3 MeV).
   For modes with 
   $\dbarpstar \to \dbarp \pi^0$ and 
   $\dbarpstar \to \dbarp \gamma$ the mass difference $\Delta M$ between
   the $\dbarpstar~$ and the $\dbarp$ must be within 3.5 MeV
   (3.5$\sigma$) and 13 MeV (2$\sigma$),
   respectively, of the expectation for $\dbarpstar$
   decays (142.2 MeV).
   
   A major background arises from $DK$ and $D^*K$ decays in which the 
   $K$ and $\pi$ in the $D$ decay are misidentified as $\pi$ and
   $K$, respectively. When this happens, the decay could be reconstructed
   as a $\Dbar K$ or $\Dstarb K$ signal event.
   To eliminate this 
   background, we recompute the invariant mass ($M_{{\rm switch}}$) 
   of the $h^+h^-$ 
   pair in $\dbarp \to h^+h^-$ switching the particle identification
   assumptions ($\pi$ vs. $K$) on the $h^+$ and the $h^-$.  We veto
   any candidates with $M_{{\rm switch}}$ within 18 MeV of the 
   known $D$ mass~\cite{PDG}; this requirement is 90\% efficient 
   on signal decays.
   In the case of $\dbarp K$, we also veto any candidate for which the
   $\dbarp$ is consistent with $D^{*+} \to D^0 \pi^+$ or 
   $D^{*0} \to D^0 \pi^0$ decay.

   \subsection{Neural Network}
   \label{sec:nn}
   After these initial selection criteria, backgrounds are overwhelmingly
   from continuum events, especially $e^+ e^- \to c \bar{c}$, with 
   $\bar{c} \to \Dzb X$, $\Dzb \to K^+ \pi^-$ and $c \to D X$,
   $D \to K^- Y$.

   The continuum background is reduced using a neural network technique.
   The neural network algorithms used for the modes with and 
   without a $D^*$  
   are slightly different.  First, for all modes we use a common neural 
   network ($NN$) based on nine quantities, listed below,
   that distinguish between 
   continuum and $B\Bbar$ events.  Then, for the modes with a $D^*$ 
   we also take advantage of the fact that the signal is distributed
   as $\cos^2\theta_{D^*}$ for $D^* \to D\pi$ or 
   $\sin^2\theta_{D^*}$ for $D^* \to D\gamma$, while the background
   is roughly independent of $\cos\theta_{D^*}$. 
%
%
   Here $\theta_{D^*}$ is the decay angle of the $D^*$, {\it i.e.},
   the angle between the direction of the $D$ and the line of flight of
   the $D^*$ relative to the parent $B$, evaluated in the $D^*$ rest
   frame.
   Thus, we
   construct a second neural network, $NN'$, which takes as 
   inputs the output of $NN$ and the value of $\cos\theta_{D^*}$.
   We then use as a selection requirement the output of $NN$ in the 
   $\dbarp K$ analysis and the output of $NN'$ in the $\dbarpstar K$ analysis.

   The nine variables used in defining $NN$ are the following:
   \begin{enumerate}
   \item A Fisher discriminant~\cite{Fisher} constructed from the 
   quantities $L_0 = \sum_i{p_i}$ and $L_2 = \sum_i{p_i \cos^2\theta_i}$
   calculated in the CM frame.  Here, $p_i$ is the momentum and
   $\theta_i$ the angle with respect to the thrust axis of the $B$ candidate
   of tracks and clusters not used to reconstruct the $B$ meson.
   \item  $|\cos \theta_T|$, where $\theta_T$ is the angle in
   the CM frame between the thrust axes of the $B$ candidate and
   the detected remainder of the event.  The distribution of 
   $|\cos \theta_T|$ is approximately flat for signal and strongly
   peaked at one for continuum background.
   \item $\cos \theta^*_B$, where $\theta^*_B$ is the polar angle
   of the $B$ candidate in the CM frame.  In this variable, the 
   signal follows a $\sin^2\theta^*_B$ distribution, while the 
   background is approximately uniform.
   \item $\cos \theta_D^K$ where $\theta_D^K$ is the decay angle
   in $\dbarp \to K\pi$.
   \item $\cos \theta_B^{D^{(*)}}$, where $\theta_B^{D^{(*)}}$
   is the decay angle
   in $B \to \dbarp K$ or
   $B \to \dbarpstar K$.  In signal events the distributions of 
   $\cos \theta_D^K$ and $\cos \theta_B^{D^{(*)}}$ are uniform. 
   On the other hand, 
   the corresponding distributions in 
   combinatorial background events tend to show
   accumulations of events near the extreme values.
   \item The charge difference $\Delta Q$ between the sum
   of the charges of tracks in the $\dbarp$ or $\dbarpstar$ hemisphere
   and the sum of the charges of the tracks in the opposite
   hemisphere excluding those tracks used in the reconstructed $B$.
   The partitioning of the event in the two hemispheres is done
   in the CM frame.
   For signal, $\langle \Delta Q \rangle = 0$, whereas
   for the $c\bar{c}$ background 
   $\langle \Delta Q \rangle \approx \frac{7}{3}\times Q_B$,
   where $Q_B$ is the charge of the $B$ candidate.  
   The value of $\langle \Delta Q \rangle $ in $c\bar{c}$ events
   is a consequence of the correlation between the charge of the
   $c$ (or $\bar{c}$) in a given hemisphere and the sum
   of the charges of all particles in that hemisphere.
   Since the
   $\Delta Q$ RMS is 2.4, this variable provides 
   approximately a $1\sigma$ separation between signal
   and $c\bar{c}$ background.
   \item $Q_B \cdot Q_K$, where $Q_K$ is the sum of the charges of all
   kaons not in the reconstructed $B$, and $Q_B$, as defined
   above, is the charge of the reconstructed $B$ candidate.
   In many signal events, there is a charged kaon among the decay 
   products of the other $B$ in the event.  The charge
   of this kaon tends to be highly correlated with the charge of the 
   $B$.  Thus, signal events tend to have $Q_B \cdot Q_K \leq -1$.  On
   the other hand, most continuum events have no kaons outside of the
   reconstructed $B$, and therefore $Q_K = 0$.
   \item The distance of closest approach between the bachelor
   track and the trajectory of the $\dbarp$.  This is
   consistent with zero for signal events, but can be
   larger in $c\bar{c}$ events.
   \item A quantity ${\cal M}_{K\ell}$ defined to be
   zero if there are no leptons ($e$ or $\mu$) in the event,
   one if there is a lepton in the event and the invariant mass
   ($m_{K\ell}$)
   of this lepton and the bachelor kaon is less than the mass of the
   $D$-meson ($m_D$), and two if $m_{K\ell} > m_D$.  This quantity
   differentiates between continuum and signal events because
   the probability of finding a lepton in a continuum event
   is smaller than in a $B\Bbar$ event.
   Furthermore, a large fraction of leptons in $c\bar{c}$ events
   are from $D \to K \ell \nu$, where $K$ is reconstructed as the 
   bachelor kaon.  For these events $m_{K\ell} < m_D$, while
   in signal events the expected distribution of $m_{K\ell}$
   extends to larger values.
\end{enumerate}

   The neural networks ($NN$ and $NN'$) are trained with simulated continuum 
   and signal events.  
   The distributions of the $NN$ and $NN'$ outputs for the control samples  
   ($D\pi$, $D^*\pi$, and off resonance data) are compared with
   expectations from the Monte Carlo simulation
   in Fig.~\ref{fig:nnplots}.  The agreement
   is satisfactory.
   We have also examined the distributions of all variables used in $NN$ and $NN'$,
   and found good agreement between the simulation and the data 
   control samples.

   Our final event selection requirement is $NN > 0.5$ for $\Dbar K$ 
   and $NN' > 0.5$ for $\Dstarb K$.  In addition, to reduce the remaining
   $B\Bbar$ backgrounds, we also require
   $\cos \theta_D^K > -0.75$.
   These final requirements are about 40\% efficient
   on simulated signal events, and reject 98.5\% of the continuum background.

   The overall reconstruction efficiencies, estimated from Monte Carlo 
   simulation, are about 14\% for $\Dbar K$,
   8\% for $\Dstarb K$ with $\Dstarb \to \Dbar \pi^0$ and
   7\% for $\Dstarb K$ with $\Dstarb \to \Dbar \gamma$.
   Note that a precise knowledge of the efficiencies is not needed
   in the analysis.
   We apply the identical
   requirements to the normalization modes $DK$ and $D^{*}K$.  Then, in the
   extraction of ${\cal R}_{K\pi}$ and ${\cal R}_{K\pi}^*$, the 
   efficiencies of the overall selection cancel in the ratio.

   \begin{table*}[bt]
   \caption{Summary of fit results.}
   \begin{center}
   \begin{tabular*}{0.85\textwidth}{@{\extracolsep{\fill}}|l|l|l|l|} \hline
        &          &  & \\
   Mode & $\Dbar K$  &  $\Dstarb K$, $\Dstarb \to \Dbar \pi^0$ & $\Dstarb K$, $\Dstarb \to \Dbar \gamma$ \\ \hline 
   Ratio of rates, ${\cal R}_{K\pi}$ or ${\cal R}^*_{K\pi}$, $\times 10^{-3}$&
   ${\cal R}_{K\pi} = 13^{+11}_{-9}$ &  
   ${\cal R}^*_{K\pi} = -2^{+10}_{-6}$ &  
   ${\cal R}^*_{K\pi} = 11^{+18}_{-13}$ \\
   Number of signal events & $5^{+4}_{-3}$ & $-0.2^{+1.3}_{-0.7}$ & $1.2^{+2.1}_{-1.4}$ \\
   Number of normalization events & $368 \pm 26$ & $150 \pm 17$ & $108 \pm 14$ \\ 
   Number of peaking charmless events                 & $0.7^{+1.4}_{-0.7}$   & $0.0^{+0.3}_{-0.0}$ & $0.1^{+0.8}_{-0.1}$ \\
   Number of peaking $D^{(*)}\pi$ events in signal sample  & $0.48 \pm 0.05$ & $0.18 \pm 0.03$ & $0.01 \pm 0.02$ \\
   \hline
   \end{tabular*}
   \end{center}
   \label{tab:fit-results}
   \end{table*}

    \begin{figure*}[hbt]
    \begin{center}
 \includegraphics[width=5.5in]{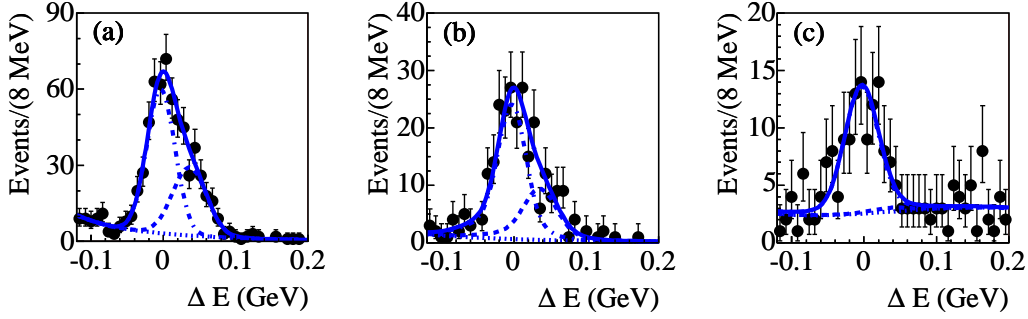}
    \caption{ $\Delta E$ distributions for normalization events
    ($DK$ and $D^*K$) with $\mes$ within 3$\sigma$ of $m_{B}$
    with the fit model overlaid.  
    (a) $DK$ events.
    (b) $D^*K$ events with $D^* \to D\pi^0$. 
    (c) $D^*K$ events with $D^* \to D\gamma$.
    The dashed (dot-dashed) lines are the contributions from $D\pi$ or $D^*\pi$
    ($DK$ or $D^*K$) 
    events.  The dotted lines are the contributions from other
    backgrounds, and the solid line is the total.}
    \label{fig:de_all}
    \end{center}
  \end{figure*}

    \begin{figure*}[hbt]
    \begin{center}
    \includegraphics[width=5.5in]{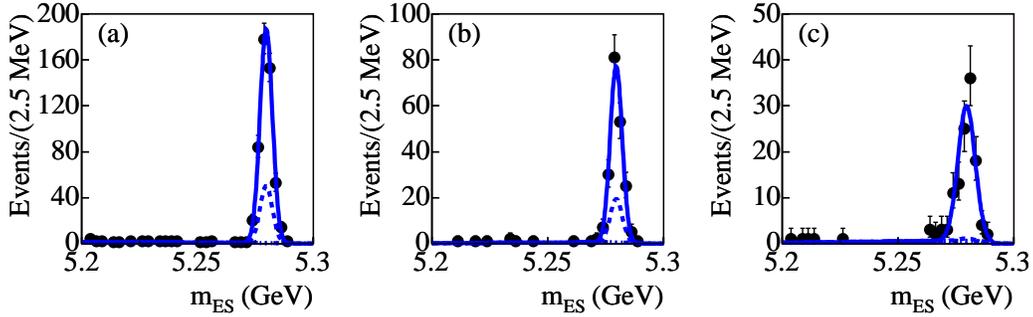}
    \caption{\protect $\mes$ distributions for normalization events
    ($DK$ and $D^*K$) with $\Delta E$ in the signal region
    with the fit model overlaid.  (a) $DK$ events.
    (b) $D^*K$ events with $D^* \to D\pi^0$. 
    (c) $D^*K$ events with $D^* \to D\gamma$.
    The dashed lines represent the backgrounds; these are mostly
    from $D\pi$ or $D^*\pi$, and also peak at the $B$-mass.  
    As explained in the text, the size of the $D\pi$ and
    $D^*\pi$ backgrounds is constrained by the simultaneous 
    fits to the distributions of Fig.~\ref{fig:de_all}.}
    \label{fig:mes_d0k_all}
    \end{center}
  \end{figure*}

    \begin{figure*}[hbt]
    \begin{center}
    \includegraphics[width=5.5in]{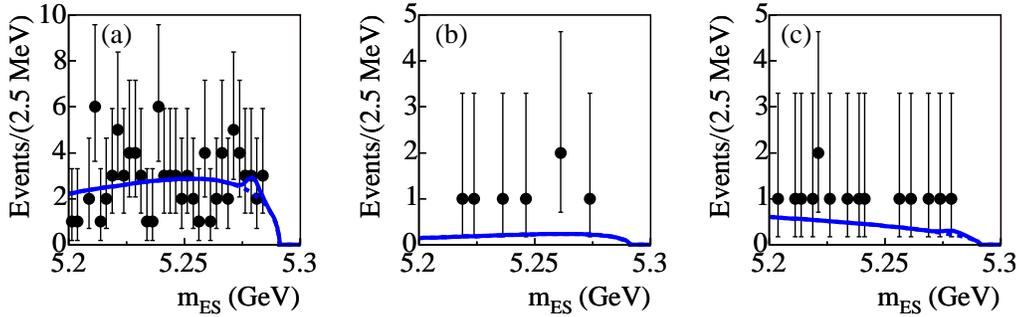}
    \caption{ $\mes$ distributions for $\Dbar K$ and $\Dstarb K$
    events with $K\pi$ mass in a sideband of the reconstructed 
    $D$ mass and
    with $\Delta E$ in the signal region.  These events are used 
    to constrain the size of possible peaking backgrounds from
    charmless
    $B$-meson decays, {\em i.e.}, decays without a $D$ in the final state.
    The fit model is overlaid.  (a) $\Dbar K$ events.
    (b) $\Dstarb K$ events with $D^* \to D\pi^0$. 
    (c) $\Dstarb K$ events with $D^* \to D\gamma$.
    Note that the $K\pi$ mass range in the sideband selection
    is a factor of 2.7 larger than in the signal selection.}
    \label{fig:mes_d0sb_all}
    \end{center}
    \end{figure*}

    \begin{figure*}[hbt]
    \begin{center}
    \includegraphics[width=5.5in]{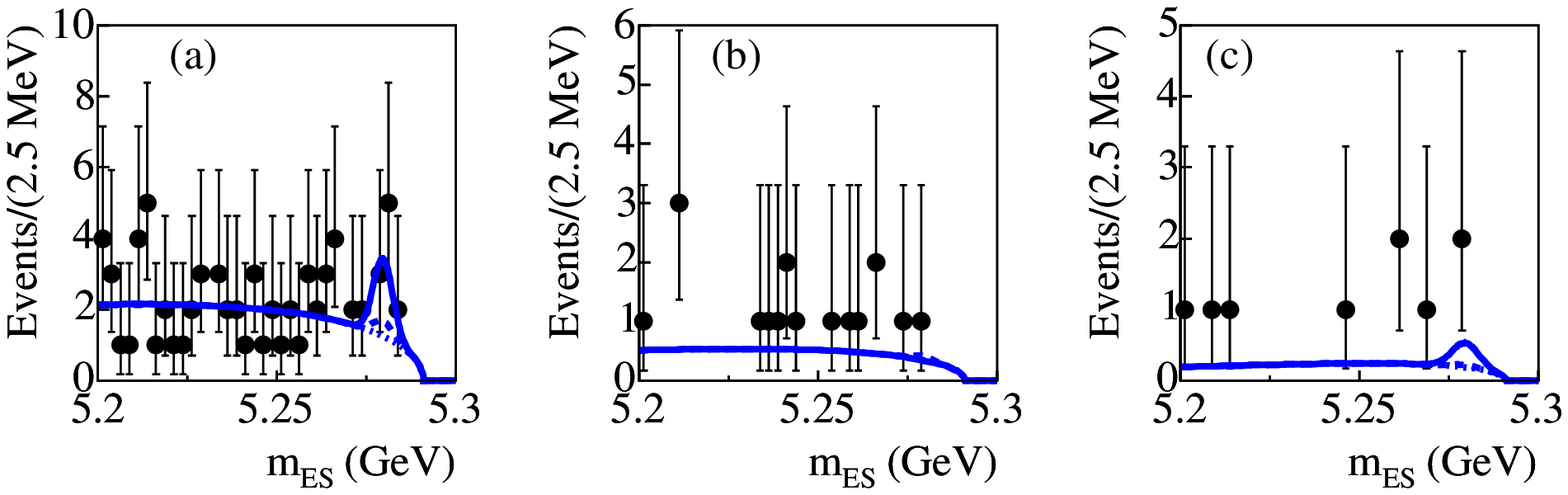}
    \caption{ $\mes$ distributions for candidate signal
    events
    with the fit model overlaid.  (a) $\Dbar K$ events.
    (b) $\Dstarb K$ events with $D^* \to D\pi^0$. 
    (c) $\Dstarb K$ events with $D^* \to D\gamma$.}
    \label{fig:mes_sig_all}
    \end{center}
    \end{figure*}

   \subsection{Fitting for event yields and 
   {\boldmath ${\cal R}^{(*)}_{K\pi}$}}
   \label{sec:fit}
   The ratios ${\cal R}_{K\pi}$ and ${\cal R}^*_{K\pi}$ are extracted
   from the ratios of the event yields in the $m_{ES}$ distribution
   for the signal modes ($\Dbar K$ and $\Dstarb K$) and the
   normalization modes ($DK$ and $\Dstar K$), taking into
   account potential differences in efficiencies and backgrounds.  All
   events must satisfy the requirements discussed above and
   have a $\Delta E$ value consistent with zero within the resolution
   ($-52~{\rm MeV} < \Delta E < 44~{\rm MeV}$). 
   Here we discuss the procedure to extract ${\cal R}_{K\pi}$;
   the values of ${\cal R}^{*}_{K\pi,D\piz}$ 
   and ${\cal R}^*_{K\pi,D\gamma}$ are obtained in the same way.

   The $\mes$ distributions for $\Dbar K$ (signal mode) and $D K$
   (normalization mode) are fitted simultaneously.  
   The fit parameter ${\cal R}_{K\pi}$ is given by 
   ${\cal R}_{K\pi} \equiv c \cdot N_{\Dbar K}/N_{DK}$,
   where $N_{\Dbar K}$ and $N_{DK}$ are the fitted yields of
   $\Dbar K$ and $DK$ events, and $c$ is a correction factor, determined
   from Monte Carlo, for the ratio of efficiencies between the two 
   modes.  We find that this factor $c$ is consistent with unity
   within the statistical accuracy of the simulation,
   $c = 0.98 \pm 0.04$ (these correction factors are 
   $c = 0.97 \pm 0.05$ and $c = 0.99 \pm 0.05$ for $D^* \to D \pi^0$ 
   and $D^* \to D \gamma$, respectively).

   The $\mes$ distributions are modeled as the sum of a threshold
   combinatorial background function~\cite{ARGUS} and a Gaussian lineshape
   centered at $m_B$.  The parameters of the background function for
   the signal mode are constrained by a simultaneous fit of the $\mes$
   distribution for events in the sideband of $\Delta E$ 
   ($-120~{\rm MeV} < \Delta E < 200~{\rm MeV}$, excluding the
   $\Delta E$ signal region defined above).
   The parameters of the Gaussian for the signal
   and normalization modes are taken to be identical. The number
   of events in the Gaussian is $N_{sig} + N_{peak}$, where $N_{sig} =
   N_{DK}$ or $N_{\Dbar K}$ and $N_{peak}$ is the number of background
   events expected to be distributed in the same way as $DK$ or
   $\Dbar K$ in $\mes$ (``peaking backgrounds'').

   There are two classes of peaking background events:

   \begin{enumerate}  
   \item Charmless $B$ decays, {\em e.g.}, $B^- \to K^+ K^- \pi^+$.
   These are indistinguishable from the $\Dbar K$ signal if the $K^-\pi^+$
   pair happens to be consistent with the $D$ mass.   
   \item Events
   of the type $B^- \to D^0 \pi^-$, where the bachelor $\pi^-$
   is misidentified as a $K^-$.  When the $D^0$ decays into
   $K^-\pi^+$ ($K^+\pi^-$), these events are indistinguishable
   in $\mes$ from $DK$ ($\Dbar K$), since $\mes$ is independent
   of particle identification assumptions.
   \end{enumerate}

   The amount of peaking 
   charmless $B$ background (1) is estimated directly from the
   data by performing a simultaneous fit to events in the sideband of
   the reconstructed $D$ mass.  In this fit the number of charmless
   background events is constrained to be $\geq 0$.

   The $\Delta E$ distribution of the $D\pi$ background (2) is 
   shifted by about $+ 50$ MeV due to the
   misidentification of the bachelor $\pi$ as a $K$. Since the
   $\Delta E$ resolution is of order 20 MeV, the $\Delta E$
   requirement does not eliminate this
   background completely.  The remaining $D\pi$ background after the
   $\Delta E$ requirement is estimated relaxing the
   $\Delta E$ requirement and performing a fit to
   the $\Delta E$ distribution of the $DK$ candidate sample, as described
   below.

   We fit the $\Delta E$ distribution of $DK$ candidates, with $\mes$
   within $3 \sigma$ of $m_B$, to the sum of a $DK$ component, a
   $D\pi$ background component, and a combinatorial background
   component, see Fig.~\ref{fig:de_all}. 
   From this fit we can estimate the number
   of $D\pi$ background events 
   after the $\Delta E$ requirement, which we denote as $N^{\pi}_{DK}$.
   In this fit, the $\Delta E$ shapes of the $DK$ and $D\pi$ components
   are constrained from the data as follows:
   \begin{itemize}
   \item The large $D\pi$
   sample, with the bachelor track identified as a pion, is used to
   constrain the shape of the $DK$ component in the sample of $DK$
   candidates.
   \item The
   same sample of $D\pi$ events, but
   reconstructed as $DK$ events, is used to
   constrain the shape of the $D\pi$ background in the $DK$ sample.
   \end{itemize}

   The $D\pi$ peaking background is much more
   important in the $DK$ (normalization) channel than in the $\Dbar K$
   (signal) channel.  This is because in order to contribute to the
   signal channel, the $D^0$ has to decay into $K^+ \pi^-$, and this
   mode is doubly CKM suppressed.
   For the $\Dbar K$ (signal) sample, the contribution from the residual 
   $D\pi$ peaking background in the $\mes$ fit is estimated as 
   $N^{\pi}_{\Dbar K} = r^2_D N^{\pi}_{DK}$, where $r_D = 0.060 \pm 0.002$
   is the ratio of the doubly CKM-suppressed to the CKM-favored
   $D \to K\pi$ amplitudes and $N^{\pi}_{DK}$ was defined above.

   The complete procedure simultaneously fits seven distributions: the
   $m_{ES}$ distributions of $DK$ and $\Dbar K$, the $\Dbar K$
   distributions in sidebands of $\Delta E$ and $m(D^0)$, the $\Delta
   E$ distribution of $DK$, and the $\Delta E$ distributions of $D \pi$
   reconstructed as $D \pi$ and as $DK$.  All fits are unbinned extended
   maximum likelihood fits.  They
   are configured in
   such a way that ${\cal R}_{K\pi}$ and ${\cal R}^*_{K\pi}$ are
   explicit fit parameters.  The advantage of this approach is that
   all uncertainties, including the uncertainties in the PDFs and the
   uncertainties in the background subtractions, are 
   correctly propagated in the statistical uncertainty reported by the
   fit. 

   The fit is performed separately for $\Dbar K$, $\Dstarb K$,
   $\Dstarb \to \Dbar \pi^0$, and $\Dstarb K$, $\Dstarb \to \Dbar
   \gamma$ and is identical for all three modes, except in the choice of
   parameterization for some signal and background components in the
   $\Delta E$ fits.   

   Systematic uncertainties in the detector
   efficiency cancel in the ratio.  
   This cancellation has been verified by studies of simulated events,
   with a statistical precision of a few percent.  
   The likelihood includes a Gaussian uncertainty term for this cancellation
   which is set by the statistical accuracy of the simulation.  Other
   systematic uncertainties, e.g., the uncertainty in the parameter
   $r_D$ used to estimate the amount of peaking backgrounds
   from $D^{(*)}\pi$, are also included in the formulation of the
   likelihood.

   The fit procedure has been extensively tested on sets
   of simulated events. It was found to provide an unbiased estimation
   of the parameters ${\cal R}_{K\pi}$ and ${\cal R}^*_{K\pi}$.

\section{RESULTS}
\label{sec:results}
  The results of the fits are displayed in Table~\ref{tab:fit-results}
  and
  Figs.~\ref{fig:de_all},~\ref{fig:mes_d0k_all},~\ref{fig:mes_d0sb_all},
  and~\ref{fig:mes_sig_all}.  As is apparent from
  Fig.~\ref{fig:mes_sig_all}, we see no evidence for the $\Dstarb K$
  and $\Dbar K$ modes.

  For the $\Dbar K$ mode we find ${\cal R}_{K\pi} = (13^{+11}_{-9})
  \times 10^{-3}$; for the $\Dstarb K$ mode we find ${\cal R}^*_{K\pi,D\piz}
  =( -2^{+10}_{-6}) \times 10^{-3}$ (for $D^* \to D \pi^0$) and ${\cal
  R}^*_{K\pi, D\gamma} = (11^{+18}_{-13}) \times 10^{-3}$ (for $D^* \to D
  \gamma$).

  Next, we use our measurements to extract information on $r_B$ and $r^*_B$.
  In the case of decays into $\dbarpstar$ we start from 
  equations~\ref{eq:star-pi} and~\ref{eq:star-gam}
  to derive
   \begin{equation}
   \label{eq:special}
   r_B^{*2} = 
   \frac{{\cal R}^{*}_{K\pi,D \piz}+{\cal R}^{*}_{K\pi,D\gamma}}{2} - r_D^2.
   \end{equation}
   We use the relationship given by equation~\ref{eq:special} in
   conjunction with $r_D = 0.060 \pm 0.002$ and
   our results for ${\cal R}^*_{K\pi,D\piz}$
   and ${\cal R}^*_{K\pi,D\gamma}$ to extract information on
   $r_B^{*2}$ with no assumptions on the values of $\gamma$ and strong phases.

\begin{figure}[hbt]
    \begin{center}
    \epsfig{file=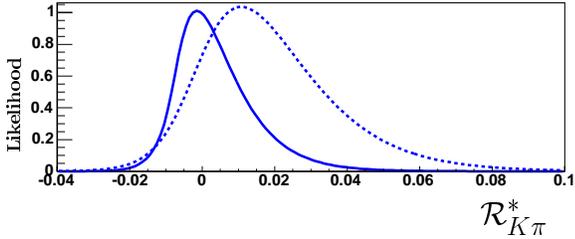,width=\linewidth}
    \caption{Likelihood functions as obtained from the
    fit described in the text for ${\cal R}^{*}_{K\pi,D\piz}$ 
    (solid line) and ${\cal R}^{*}_{K\pi,D\gamma}$ (dashed line).} 
    \label{fig:rk-star}
    \end{center}
    \end{figure}

\begin{figure}[hbt]
    \begin{center}
    \epsfig{file=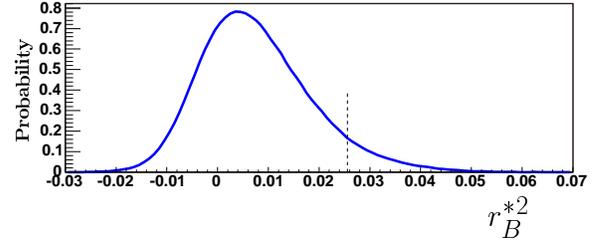,width=\linewidth}
    \caption{Probability distribution function (arbitrary units)
    for $r_B^{*2}$ obtained
    as described in the text. 
    The integral of the 
    function for $0 < r_B^{*2} < (0.16)^2$ is 
    nine-tenths of the integral for $r_B^{*2}>0$.
    The vertical dashed line is drawn at $r_B^{*2} = (0.16)^2$.}
    \label{fig:likelihood2}
    \end{center}
    \end{figure}

   Since the uncertainties in ${\cal R}^{*}_{K\pi,D\piz}$
   and ${\cal R}^{*}_{K\pi,D\gamma}$ are non-Gaussian, care must
   be taken in propagating them into an uncertainty in $r_B^{*2}$.
   We interpret the fit likelihoods for ${\cal R}^{*}_{K\pi,D\piz}$
   and ${\cal R}^{*}_{K\pi,D\gamma}$ (see Figure~\ref{fig:rk-star})
   as posterior PDFs assuming constant priors.  We assume a Gaussian
   PDF for $r_D$.  We then convolve numerically the three PDFs 
   for ${\cal R}^{*}_{K\pi,D\piz}$, ${\cal R}^{*}_{K\pi,D\gamma}$, and
   $r_D$ according to equation~\ref{eq:special} to obtain a
   PDF for $r_B^{*2}$ which is shown in Figure~\ref{fig:likelihood2}.
   The convolution relies on the fact that
   the measurements of ${\cal R}^{*}_{K\pi,D\piz}$
   and ${\cal R}^{*}_{K\pi,D\gamma}$ are uncorrelated
   (the correlation due to the uncertainty in $r_D$, which was
   used to extract the $D\pi$ peaking backgrounds in the two
   modes, is negligible).

   Our result is $r_B^{*2} = (4^{+14}_{-8}) \times 10^{-3}$.
   Based on the PDF for $r_B^{*2}$ shown in 
   Figure~\ref{fig:likelihood2}
   we set an upper limit $r_B^{*2} < (0.16)^2$ at the 90\% C.L. 
   using a Bayesian method
   with a uniform prior for $r_B^{*2} > 0$.

    \begin{figure}[thb]
    \begin{center}
    \epsfig{file=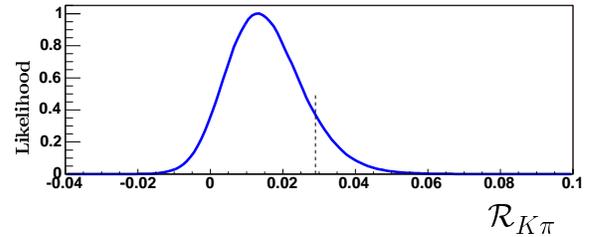,width=\linewidth}
    \caption{ Likelihood function (arbitrary units)
    for ${\cal R}_{K\pi}$ as extracted from the fit described
    in the text.  The integral of the likelihood 
    function for $0 < {\cal R}_{K\pi} < 0.029$ is 
    nine-tenths of the integral for ${\cal R}_{K\pi} >0$.
    The vertical dashed line is drawn at ${\cal R}_{K\pi} = 0.029$.}
    \label{fig:likelihood}
    \end{center}
    \end{figure}

\begin{figure}[htb]
   \begin{center}
   \epsfig{file=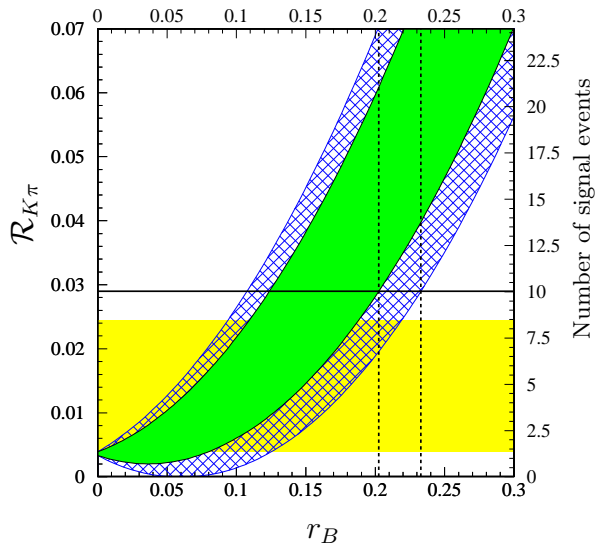,width=\linewidth}
   \caption{Expectations for ${\cal R}_{K\pi}$ and the number
   of signal events {\it vs.} $r_B$.  Dark filled-in area:
   allowed region for any value of $\delta$, with a $\pm
   1\sigma$ variation on $r_D$, and $51^{\circ} < \gamma <
   66^{\circ}$.  Hatched area: additional allowed region with no
   constraint on $\gamma$.  Note that the uncertainty on $r_D$ has a
   very small effect on the size of the allowed regions.  The
   horizontal line represents the 90\% C.L. limit ${\cal R}_{K\pi} <
   0.029$. The vertical dashed lines
   are drawn at $r_B = 0.203$ and $r_B = 0.233$.  
   They represent the 90\% C.L. upper limits on
   $r_B$ with and without the constraint on $\gamma$. The
   light filled-in area represents the 68\% C.L. region corresponding
   to ${\cal R}_{K\pi} = 0.013 \pm^{0.011}_{0.009}$. }
   \label{fig:ads_rate}
   \end{center}
   \end{figure}

  In the case of decays into a $D/\Dbar$, there is not enough information to 
  extract the ratio $r_B$ without additional assumptions.  Thus, we first extract an 
  upper limit on the experimentally measured
  quantity ${\cal R}_{K\pi}$.  This is done starting 
  from the likelihood as a function of ${\cal R}_{K\pi}$ 
  (see Fig.~\ref{fig:likelihood})
  using a Bayesian method
  with a uniform prior for ${\cal R}_{K\pi} > 0$.
  The limit is ${\cal R}_{K\pi} < 0.029$
  at 90\%C.L.
   Next, in Fig.~\ref{fig:ads_rate} we show the dependence of
   ${\cal R}_{K\pi}$ on $r_B$, together with our limit on 
   ${\cal R}_{K\pi}$.
   This dependence is shown allowing a $\pm 1\sigma$ variation on $r_D$,
   for the full range $0^{\circ}-180^{\circ}$ for $\gamma$ and
   $\delta$, as well as with the restriction
   $51^{\circ} < \gamma < 66^{\circ}$ suggested by global CKM
   fits~\cite{ckmfitter}.  We use the information displayed
   in this Figure to set an upper limit on $r_B$. 
   The least restrictive limit on $r_B$
   is computed assuming maximal destructive interference
   between the $b \to c$ and $b \to u$ amplitudes:
   $\gamma=0^{\circ}, \delta = 180^{\circ}$ or 
   $\gamma=180^{\circ}, \delta = 0^{\circ}$.  
   The limit is $r_B < 0.23$ at 90\% C.L.


   \begin{table}[hbt]
   \caption{Summary of the results of this analysis.  }
   \begin{center}
   \begin{tabular}{|c|c|c|} \hline
        & Measured Value   &  90\% C.L. limit  \\ \hline
${\cal R}_{K\pi}$          & $ 0.013 \pm^{0.011}_{0.009}$ & $< 0.029$ \\
${\cal R}^*_{K\pi,D\piz}$  & $-0.002 \pm^{0.010}_{0.006}$ & $< 0.023$ \\
${\cal R}^*_{K\pi,D\gamma}$& $ 0.011 \pm^{0.018}_{0.013}$ & $< 0.045$ \\
$r_B$                      & ...                         & $< 0.23$ \\
$r^{*2}_B$                    & $0.004 \pm^{0.014}_{0.008}$   & $< (0.16)^2$ \\ \hline
   \end{tabular}
   \end{center}
   \label{tab:all-results}
   \end{table}

\section{SUMMARY}
\label{sec:Summary}

   In summary, we find no significant evidence for the decays $B^{\pm}
   \to [K^{\mp}\pi^{\pm}]_D K^{\pm}$ and 
   $B^{\pm} \to [K^{\mp}\pi^{\pm}]_{D^*} K^{\pm}$.  
   We set upper limits on the 
   ratios ${\cal R}^{(*)}_{K\pi}$ of the rates for these modes 
   and the favored modes $B^{\pm} \to [K^{\pm}\pi^{\mp}]_D K^{\pm}$ 
   and $B^{\pm} \to [K^{\pm}\pi^{\mp}]_{D^*} K^{\pm}$.
   We also use our data to set upper limits on the ratios of
   $b \to u$ and $b \to c$ amplitudes $r_B$ and $r^*_B$.
   All of our results are summarized in Table~\ref{tab:all-results}.

   Our results favor values of $r_B$ and $r_B^*$ somewhat smaller than the
   value of 0.2 which can be estimated from the measurements of
   color-suppressed $b \to c$ transitions~\cite{gronau}.
   If $r_B$ and $r_B^*$ are
   small, the suppression of the
   $b\rightarrow u$ amplitude will make the determination of $\gamma$
   using methods based on the interference of the diagrams in
   Fig.~\ref{fig:feynman} difficult.

\section{ACKNOWLEDGMENTS}
\label{sec:Acknowledgments}

We are grateful for the 
extraordinary contributions of our \pep2\ colleagues in
achieving the excellent luminosity and machine conditions
that have made this work possible.
The success of this project also relies critically on the 
expertise and dedication of the computing organizations that 
support \babar.
The collaborating institutions wish to thank 
SLAC for its support and the kind hospitality extended to them. 
This work is supported by the
US Department of Energy
and National Science Foundation, the
Natural Sciences and Engineering Research Council (Canada),
Institute of High Energy Physics (China), the
Commissariat \`a l'Energie Atomique and
Institut National de Physique Nucl\'eaire et de Physique des Particules
(France), the
Bundesministerium f\"ur Bildung und Forschung and
Deutsche Forschungsgemeinschaft
(Germany), the
Istituto Nazionale di Fisica Nucleare (Italy),
the Foundation for Fundamental Research on Matter (The Netherlands),
the Research Council of Norway, the
Ministry of Science and Technology of the Russian Federation, and the
Particle Physics and Astronomy Research Council (United Kingdom). 
Individuals have received support from 
CONACyT (Mexico),
the A. P. Sloan Foundation, 
the Research Corporation,
and the Alexander von Humboldt Foundation.

\end{document}